\documentclass[intlimits,twoside,a4paper]{article}

\usepackage{color}
\usepackage{amsmath,amssymb}
\usepackage{graphicx}

\usepackage[T2A]{fontenc}
\usepackage[cp1251]{inputenc}

\usepackage[eqsecnum]{cmpj2}

\issue{2013}{16}{4}{43604}
\doinumber{10.5488/CMP.16.43604}

\title[Consistency of water potentials with diffraction data]%
{Comparison of the TIP4P-2005, SWM4-DP and BK3 interaction potentials of liquid water with respect to their consistency with neutron and X-ray diffraction data of pure water\thanks{Dedicated to Prof. Dr. Myroslav Holovko on the occasion of his 70$^\mathrm{th}$ birthday}}

\author[Z. Steinczinger, L. Pusztai]{Z. Steinczinger\refaddr{label1},
        L. Pusztai\refaddr{label2}\thanks{pusztai.laszlo@wigner.mta.hu}}
\addresses{
\addr{label1} Budai Nagy Antal Secondary School, H--1121, Budapest, Anna utca 13--15, Hungary
\addr{label2} Institute for Solid State Physics and Optics, Wigner Research Centre for Physics, Hungarian Academy of Sciences, H--1525 Budapest, P.O. Box 49, Hungary
}

\authorcopyright{Z. Steinczinger, L. Pusztai, 2013}
\date{Received August 12, 2013, in final form September 2, 2013}

\begin{document}

\maketitle

\begin{abstract}
Following a fairly comprehensive study on popular interaction potentials of water (Pusztai et al., J. Chem. Phys., 2008, \textbf{129}, 184103), here two more recent polarizable potential sets, SWM4-DP (Lamoureux et al., Chem. Phys. Lett., 2006, \textbf{418}, 245) and BK3 (Kiss et al., J. Chem. Phys., 2013, \textbf{138}, 204507) are compared to the  TIP4P-2005 water potential (Abascal et al., J. Chem. Phys., 2005, \textbf{123}, 234505) that had previously appeared to be most favoravble. The basis of comparison was the compatibility with the results of neutron and X-ray diffraction experiments on pure water, using the scheme applied by Pusztai et al. (2008). The scheme combines the experimental total scattering structure factors (TSSF) and partial radial distribution functions (PRDF) from molecular dynamics simulations in a single structural model. Goodness-of-fit values to the O--O, O--H and H--H simulated PRDF-s and to the experimental neutron and X-ray TSSF provided a measure that can characterize the level of consistency between interaction potentials and diffraction experiments. Among the sets of partial RDF-s investigated here, the ones corresponding to the SWM4-DP potential parameters have proven to be the most consistent with the particular diffraction results taken for the present study, by a hardly significant margin ahead of BK3. Perhaps more importantly, it is shown that the three sets of potential parameters produce nearly equivalent PRDF-s that may all be made consistent with diffraction data at a very high level. The largest differences can be detected in terms of the O--O partial radial distribution function.
\pacs 61.20.-p, 61.25.-f, 61.05.fm
\end{abstract}

\section{Introduction}

When assessing the results of computer simulations that use interaction
potentials, comparison with experimental data is essential.
For the microscopic structure the proper quantity to compare with is the total scattering structure factor (TSSF), $F(Q)$, that is related to the partial radial distribution functions (PRDF) via:
\begin{equation}
\label{GR-def}
G(r) = \sum_{i,j=1} ^{n} c_{i} c_{j} b_{i} b_{j} [g_{i,j}(r)-1],
\end{equation}
\begin{equation}
\label{FQ-def}
F(Q) = \rho_{0}\int_{0}^{\infty}4 \pi r^2 G(r)[\sin(Qr)/Q] \rd r.
\end{equation}
In the above equations, \emph{c$_{i}$} and \emph{b$_{i}$} are the molar ratio and the neutron scattering length of species \emph{i}, \emph{G(r)} is the total radial distribution function, \emph{$\rho_{0}$} is the number density and \emph{Q} is the scattering variable (proportional to the scattering angle); indices \emph{i} and \emph{j} run through nuclear species. (For details of the formalism used here, see reference~\cite{keen}).

Pure liquid water is arguably the system that has been targeted by the greatest number of computer simulation studies of all pure materials (see, e.g., works of Prof. Holovko on the subject \cite{myroslav-1,myroslav-2}), resulting in an excessive number of interaction potential parameter sets. The importance of the issue is reflected by the fact that new potential parameters have been introduced even in this very calendar year \cite{bk3}. Reviews describing many of the available interaction potential models introduced for liquid water are, for instance, references \cite{guillot,vega}. Unfortunately, the overwhelming majority of computer simulation studies show comparisons with the `experimental' PRDF-s only \cite{bk3,sw,tip}, whose functions have been shown to be only interpretations of the measured diffraction data \cite{prb99,rmcmd,rmcmd-water,zsuzsi-1}. The proper quantity to cross-check with would be the direct experimental information, the total scattering structure factors. In order to simply resolve this (mostly, technical) issue, a simple protocol was introduced some years ago \cite{rmcmd}, that was later applied to the case of pure water \cite{rmcmd-water}. The main finding of this latter investigation was, in short,  that out of the 8 water potential models considered therein, it was TIP4P-2005 \cite{tip} that proved to be consistent at the highest level with a given neutron diffraction dataset \cite{neutron} taken on pure liquid heavy water.

In this work, we wish to extend the investigation described in reference \cite{rmcmd-water}, by

\begin{enumerate}

  \item[(a)] adding two recent polarizable water potential models, SWM4-DP \cite{sw} and BK3 \cite{bk3}; the reasons for picking exactly these two potential parameter sets was that, on the one hand, SWM4-DP could be very successfully applied to structural studies of various concentrated aqueous electrolyte solutions \cite{csf,licl}, and, on the other hand, the very recent BK3 set is claimed to be capable of outperforming most water potentials in a detailed comparison with a great number of properties \cite{bk3};

  \item[(b)] scrutinizing the compatibility  not only with neutron-, but also with X-ray diffraction data.

\end{enumerate}

Here, results for SWM4-DP and BK3 will be compared to those obtained for TIP4P-2005; this is the way we wish to find a direct link to our earlier study \cite{rmcmd-water}. As experimental data, the same neutron diffraction results as considered previously are taken from reference \cite{neutron} on liquid heavy water together with  one of the most recent published sets of X-ray data of Fu et al. \cite{xray}.

\section{Reverse Monte Carlo modelling}
The protocol mentioned above \cite{rmcmd} is based on a now 25 year old technique of structural modelling, the so-called Reverse Monte Carlo (RMC) method \cite{rmc} and, therefore, a short description of the method may be appropriate here.

Reverse Monte Carlo \cite{rmc,robert-review,rmc++_jpcm,rmc++_joam,rmc_pot} is a simple tool for constructing large, three-dimensional structural models that are consistent with total scattering structure factors (within the estimated level of their errors) obtained from diffraction experiments. Via random movements of particles, the difference  between experimental and model total structure factors (calculated similarly to the \emph{$\chi^{2}$}-statistics) is minimized. As a result, by the end of the calculation, a particle configuration is available that is consistent with the experimental structure factor(s). If the structure is analyzed further, partial radial distribution functions, as well as other structural characteristics (neighbor distributions, cosine distribution of bond angles) can be calculated from the particle configurations.

A possible algorithm that can realize the above features may be outlined as follows \cite{rmc}:

\begin{enumerate}
  \item Start with an initial collection of particle coordinates in a cubic box; this may be a crystalline or a random distribution of at least a few thousand of particles, or even the final particle distribution from the previous simulation.

  \item Calculate the partial radial distribution functions for the configuration. Compose total radial distribution functions according to the experimental weighting factors. Use Fourier transformation for calculating total scattering structure factors.

  \item Calculate differences between model and experimental functions as follows (shown here for one single TSSF):
\begin{equation}
\label{CHI2-def}
\chi^{2} [F(Q)] = \sum_{i} \left[F^\textrm{C} (Q_{i}) - F^\textrm{E} (Q_{i})\right]^{2} \big/ \sigma^{2}\, .
\end{equation}
The `C' and `E' superscripts refer to `calculated' and `experimental' functions, respectively; $\sigma$ is a control parameter that is related to the assumed level of experimental errors.

  \item Move one particle at random.

  \item Calculate PRDF-s, TRDF-s, TSSF-s and, from them, also the $\chi^{2}$, for the new position.

  \item If the $\chi^{2}$ for the new position is smaller than it was for the old position (i.e., the difference between simulated and measured TSSF-s has become smaller), then accept the move immediately. Otherwise accept the move only with a probability that is proportional to $\exp(-\Delta \chi^{2})$; accepting `bad' moves with such small but finite probability will prevent calculations from sticking in local minima. If a move is `accepted' then the `new' position becomes the `old' one for the next attempted move.
  \item Continue from step 4.
\end{enumerate}

The most valuable feature of the RMC method concerning the present investigation is that it can incorporate any piece of information that can be calculated directly from particle coordinates. Partial radial distribution functions coming from computer simulations are obviously these types of data as well as the measured total structure factors. If a consistency (i.e., agreement within errors) with all input data is reached, then it may be stated that these input data are mutually consistent. On the other hand, if some of the input data cannot be approached within their uncertainties, it means that a particular part of the input data set is not consistent with the other pieces of input information. In our case, this would mean that some of the input PRDF-s from MD simulations would not be consistent with the experimental input total scattering structure factor(s).

In the RMC calculations that are the basis of the present work, one total scattering structure factor from neutron diffraction \cite{neutron}, one TSSF from X-ray diffraction \cite{xray} and three partial radial distribution functions (O--O, O--H and H--H) from references \cite{bk3,sw,tip} are used as input data for each RMC calculation. That is, in conjunction with each of the 3 potential parameter sets \cite{bk3,sw,tip} RMC is performed with the requirement that 5 input datasets (neutron TSSF, X-ray TSSF, and O--O, O--H and H--H PRDF-s from molecular dynamics simulations described in references \cite{bk3,sw,tip}) should be approached within the errors simultaneously. The primary condition assumed that experimental data must be reproduced at the same level as they are approached in the absence of PRDF-s from MD; this goal could be achieved by systematically varying the $\sigma$ control parameters (see above) for each individual dataset. On average, at least 5 independent calculations were needed to find the proper balance between individual $\sigma$ parameters; thus, the total number of calculations reported in this study is over 40 (see table~\ref{table1}).

Both experimental and simulation data were obtained at 298~K and 0.1~MPa (ambient conditions); the molecular number density was 0.0334 molecules {\AA}$^{-3}$ in each case. Molecular dynamics simulations that resulted in the PRDF-s used in this work were performed in the canonical (NVT) ensemble \cite{bk3,sw,tip} at ambient conditions. Since the number of particles and the volume remain unchanged during a RMC calculation, the NVT ensemble is a rather natural analogue for our Reverse Monte Carlo computations as well (with the notion that temperature appears only implicitly, via the experimental number density and diffraction data applied). In each Reverse Monte Carlo run, the simulation box contained 2000 water molecules (6000 atoms). Typically, `equilibration' (i.e., reaching the state where the fits to input data have not improved further and started to fluctuate) last for a few million accepted moves. Shorter refinement calculations (e.g., when small modifications of $\sigma$ parameters for only one or two datasets were made) were run for about 200000 accepted moves. Goodness-of-fit values, $R_{w}$-s (which are sums of the squared differences, see below), are reported in a normalised form, so that variations in terms of the number of \emph{r} and \emph{Q} points considered would not affect the assessment; additionally, the applied \emph{r} and \emph{Q} ranges were kept as uniform as  possible.

Definitions of the various $R_{w}$-s used throughout this contribution are provided below (for PRDF-s, only the example of the O--O $g(r)$ is shown):
\begin{equation}
\label{RWFQ-def}
{R_{w}}^2 [F(Q)] = \sum_{i} \left[F^\textrm{C} (Q_{i}) - F^\textrm{E} (Q_{i})\right]^{2} \big/ \sum_{i} F^\textrm{E} (Q_{i})^{2}\,,
\end{equation}
\begin{equation}
\label{RWgr-def}
{R_{w}}^2 [g_\textrm{OO}(r)] = \sum_{j} \left[g_\textrm{OO}^\textrm{C} (r_{j}) - g_\textrm{OO}^\textrm{E} (r_{j})\right]^{2} \big/ \sum_{j} g_\textrm{OO}^\textrm{E} (r_{j})^{2}\, ,
\end{equation}
where $N_{i}$ and $N_{j}$ are the numbers of $Q$ and $r$ points, respectively, for the experimental TSSF-s and `experimental' (MD simulated)  $g(r)$-s, respectively. Indices `C' and `E' refer to `RMC calculated' and `experimental' quantities. In table~\ref{table1} below, the square roots of the left hand side expressions are shown in the units of `\%'.

\section{Results and discussion}

\begin{figure}[h]
\centerline{
\includegraphics[angle=270, width=0.6\textwidth]{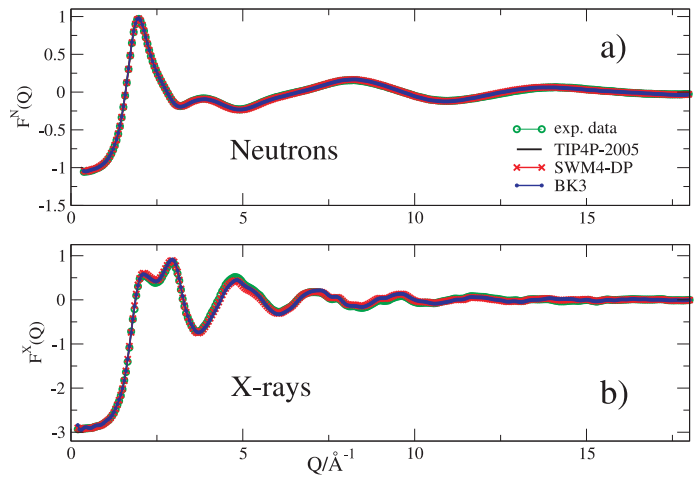}
}
\caption{(Color online) RMC modelling partial radial distribution functions of liquid water from various MD simulations \cite{bk3,sw,tip}, together with neutron diffraction TSSF of heavy water \cite{neutron} (part~a) and X-ray diffraction TSSF of light water \cite{xray} (part~b). Note that each curve runs together with the experimental results, indicating the agreement without any visible deviations from experiment.}
\label{fig1}
\end{figure}

Table \ref{table1} summarizes the main findings of the present work, namely the goodness-of-fit values for each calculation reported here. It is instructive to look at figure~\ref{fig1} in parallel, so that the level of consistency with experimental TSSF-s may be appreciated in terms of both numbers and `visual inspection'. It is clear that each PRDF-containing RMC run produced a full agreement, within the estimated error levels and without any visible sign of any disagreement, with both sets of diffraction data.

\begin{table}[ht]
\caption{Goodness-of-fit ($R_{w}$) values for individual data sets [$F^N(Q)$, $F^X(Q)$ and the three partial $g(r)$-s] for the three water potentials considered here. $R_{w}$-s for calculations without experimental data are also quoted, so that the influence of experimental data may be assessed. (`$R_{w}$  sum' is the sum of the individual $R_{w}$ values; ND: neutron diffraction; XRD: X-ray diffraction.)\label{table1}}
\vspace{2ex}
\begin{tabular}{|l||c|c|c|c|c|c|c|}
\hline
  &TIP4P-2005&TIP4P-2005&SWM4-DP&SWM4-DP&BK3&BK3&Exp\\
  &Exp+MD&MD&Exp+MD&MD&Exp+MD&MD&\\
\hline\hline
$R_{w}$[ND $F(Q)$]&2.0\% & -- & 2.2\% & -- & 2.6\% & -- & 1.2\%\\
\hline
$R_{w}$[XRD $F(Q)$]&5.2\% & -- & 4.1\%& -- & 4.9\% & -- & 4.1\% \\
\hline
$R_{w}$[$g_\textrm{OO}(r)$]& 10.0\% & 4.5\% & 11.5\% & 0.1\% & 8.4\% & 6.6\% & --\\
\hline
$R_{w}$[$g_\textrm{OH}(r)$]& 8.0\% & 4.0\% & 6.3\% & 2.0\% & 7.7\% & 4.3\% & -- \\
\hline
$R_{w}$[$g_\textrm{HH}(r)$]& 4.5\%  & 1.8\% & 4.0\% & 1.0\% & 4.7\% & 2.6\% &  -- \\
\hline
$R_{w}$ sum&29.7\%& (11.3\%) &  28.1\% & ( 3.1\%) & 28.3\% & ( 13.5\%) & -- \\
\hline
\end{tabular}
\end{table}

In table \ref{table1}, the $R_{w}$ values for the reference RMC runs, containing either PRDF-s \textit{only} or TSSF-s \textit{only}, are also quoted.
Clearly, the modelling of TSSF-s and PRDF-s together causes the deterioration of the goodness-of-fit; differences between $R_{w}$-s
obtained for `combined' (TSSF+RSDF) and `PRDF-only' calculations can give an idea as to how far from the MD results the best match to the
experimental TSSF-s lies. In this sense, SWM4-DP PRDF-s seem to suffer the largest deviation from the end results of MD simulations \cite{sw},
whereas BK3 PRDF-s \cite{bk3} should have changed the least.

The level of consistency between a given interatomic potential and the two sets of experimental TSSF-s is measured by the overall $R_{w}$ values (`$R_{w}$ sum' in table~\ref{table1}) of the `combined' Reverse Monte Carlo calculations. According to this measure, it is the SWM4-DP potential \cite{sw} that comes out as the best (lowest `$R_{w}$ sum'), but only by a very small gap ahead of BK3 \cite{bk3}. TIP4P-2005 \cite{tip} performs noticeably worse~--- but still, the lag behind the two polarizable models is quite small. Considering that TIP4P-2005 was the clear `winner' in the previous study (although using neutron diffraction data only) \cite{rmcmd-water}, the improvement of water potentials in terms of the structure is remarkable (even though only ambient pure water has been investigated here).

\begin{figure}[!b]
\centerline{
\includegraphics[angle=270, width=0.7\textwidth]{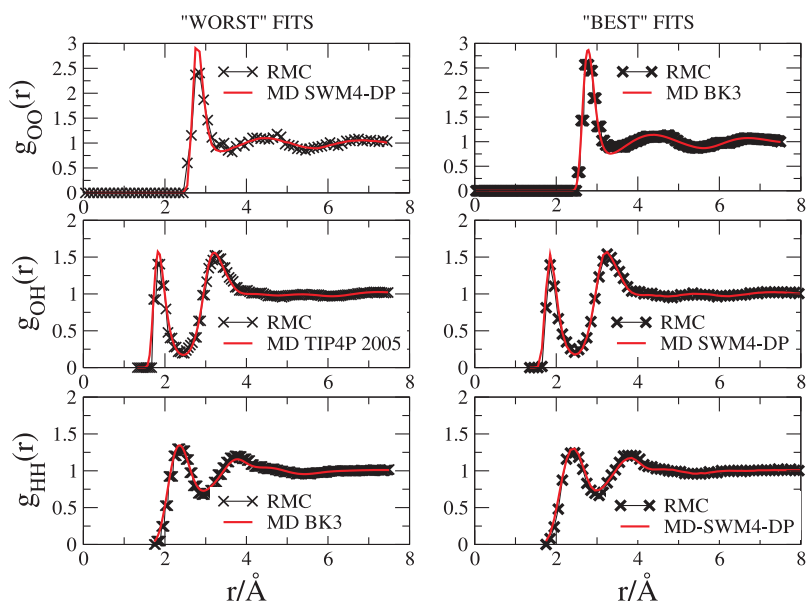}
}
\caption{(Color online) RMC modelling MD simulated O--O (upper panels), O--H (mid panels) and H--H (bottom panels) partial radial distribution functions of liquid water, together with the neutron diffraction TSSF of heavy water\cite{neutron} and X-ray diffraction TSSF of light water\cite{xray}. Left hand panels: worst cases; right hand panels: best cases. Note that differences between `worst' and `best' are hardly visible.}
\label{fig2}
\end{figure}

Figure \ref{fig2} displays comparisons of MD \cite{bk3,sw,tip} and RMC calculated O--O, O--H and H--H PRDF-s in groups of `worst' and `best' fits. Clearly, the O--O PRDF-s seem to be the hardest to be made consistent with experimental data (see also table~\ref{table1}). This is very different from the findings reported in reference \cite{rmcmd-water}, where the O--H partial appeared to be the most problematic. In this respect, the fact that in reference \cite{rmcmd-water} only one experimental neutron TSSF was used makes a huge difference: the relevant contribution (`weighting factor') of the O--O PRDF to the neutron-weighted TSSF of pure heavy water is somewhat less than 10 \%. That is, the neutron data considered \cite{neutron} are rather insensitive to the variations of the O--O PRDF. The appearance of X-ray diffraction data makes a huge difference: such data for water are sometimes evaluated so that purely the O--O contribution (in the form of a `center-of-mass TSSF' \cite{xray,narten}) is quoted as final results. That is, the TSSF from X-ray diffraction experiments poses a very strong constraint (and consequently, leaves very little freedom) for the O--O partial. Since there is still a controversy about the most appropriate way of measuring and correcting X-ray diffraction data on water (see, e.g., reference \cite{skinner}), the issue should be investigated further in more detail, involving old \cite{narten} and new \cite{skinner} results alike.

\begin{figure}[h]
\centerline{
\includegraphics[angle=270, width=0.6\textwidth]{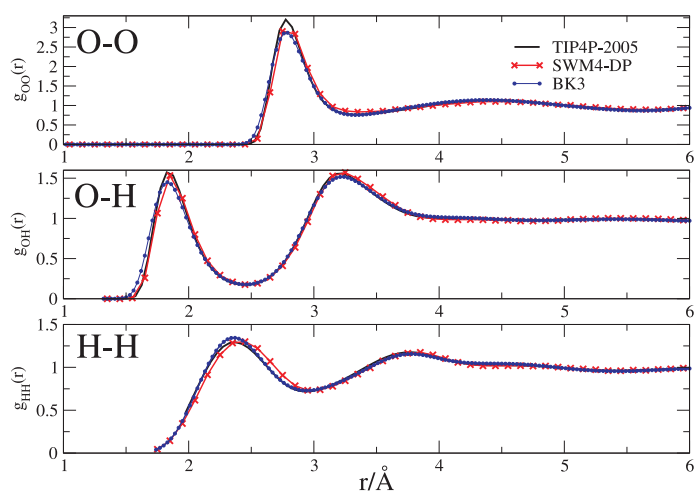}
}
\caption{(Color online) Partial radial distribution functions of MD simulated liquid water. Upper panel: $g_\textrm{OO}(r)$; mid panel: $g_\textrm{OH}(r)$; bottom panel: $g_\textrm{HH}(r)$. Note that differences between corresponding partials are much smaller than reported in reference \cite{rmcmd-water}.}
\label{fig3}
\end{figure}

Figure \ref{fig3} compares PRDF-s resulting from MD simulations of the three water potentials \cite{bk3,sw,tip} considered in this work. From a distance (even from a short one), the corresponding partials look remarkably similar to each other, much more so than reported by figure~3 of reference \cite{rmcmd-water}. Therefore, it is quite in order to diagnose that over the past few years, the two-particle level structure of ambient liquid water brought about by various modern interatomic potential parameter sets seems to be converging. However, it is still worthwhile  emphasizing some of the clear (although small) differences that may help develop the  potentials even further on. The most successful O--O PRDF (see table~\ref{table1}), from BK3, appears to be more structured than its SWM4-DP equivalent but has a clearly weaker first maximum than that from TIP4P-2005. Also note that, unlike for the other two partials, there is no dispute about the position of the main peak here. As concerns the O--H PRDF, SWM4-DP turns out to be most appropriate: this potential produces the longest H-bonding distance (which is the first intermolecular O{\ldots}H distance). This is consistent with the findings of references \cite{prb99,rmcmd-water}. Also, SWM4-DP causes the most definite oscillations at higher $r$ values. Finally, for the H--H PRDF SWM4-DP, that seems to be somewhat better than the other two potentials, brings about the longest first intermolecular H{\ldots}H distance, again, with a little better defined ordering at larger distances. In summary, it seems surprising that for all the three PRDF-s, the potential that produced the strongest long range ordering (by however small a margin) always turned out to be the best performer.

\section{Conclusions and outlook}

It has been shown that three modern interaction potential parameter sets for liquid water, TIP4P-2005 \cite{tip}, SWM4-DP \cite{sw} and BK3 \cite{bk3}, are nearly equally consistent with the measured neutron \cite{neutron} and X-ray \cite{xray} diffraction data of liquid water at ambient conditions (see figure~\ref{fig2}). Meticulous comparison (see table~\ref{table1}) reveals that a distinction may be made between the 2 polarizable potentials (SWM4-DP and BK3) and the non-polarizable one (TIP4P-2005): in terms of the structure, these up-to-date polarizable models perform somewhat better than the `champion' of the previous, analogous investigation \cite{rmcmd-water}.

The findings of the current contribution call for a more comprehensive study of interatomic potentials of water, involving a larger number of potential parameter sets and, most importantly, further recent X-ray diffraction data sets. This latter issue seems to have become outstanding recently: the latest large scale comparison between the available X-ray diffraction results \cite{skinner} shows that even during the past 5 years, the measured X-ray weighted TSSF of pure liquid water keeps changing. On the other hand, the neutron weighted TSSF of pure heavy water has not been altered over at least the recent 15 years: this fact suggests that a detailed investigation concerning the consistency between the measured neutron- and X-ray diffraction data would seem desirable, with the neutron data acting as the cornerstone TSSF.

\section*{Acknowledgements}

This work has been supported by the Hungarian Basic Research Fund (OTKA), Grant No.~K083529.


\newpage
\ukrainianpart

\title%
{Порівняння потенціалів взаємодії  TIP4P-2005, SWM4-DP і BK3 для
рідкої фази води з точки зору їх узгодження з даними по розсіянню
нейтронів і рентгенівського випромінювання для чистої води}

\author{Ж. Штеінцінгер\refaddr{label1},
        Л. Пустаї\refaddr{label2}}
\addresses{
\addr{label1} Школа другого ступеня Будаі Надь Антал, H--1121,
Будапешт, Угорщина
\addr{label2} Інститут фізики і оптики твердого
тіла, Вігнерівський дослідний центр фізики, Угорська академія наук,
H--1525 Будапешт, Угорщина }

\makeukrtitle

\begin{abstract}
\tolerance=3000%
Слідуючи досить повному вивченню популярних потенціалів взаємодії
для води (Pusztai et al., J. Chem. Phys., 2008, \textbf{129},
184103), у цій роботі два більш сучасних потенціали з врахуванням
поляризованості молекул типу  SWM4-DP (Lamoureux et al., Chem. Phys.
Lett., 2006, \textbf{418}, 245) і BK3 (Kiss et al., J. Chem. Phys.,
2013, \textbf{138}, 204507) порівняно з моделлю   TIP4P-2005
(Abascal et al., J. Chem. Phys., 2005, \textbf{123}, 234505), яка
донедавна вважалася найуспішнішою. Основою для порівняння є
узгодження з результатами експериментів по розсіянню нейтронів та
рентгенівського випромінювання для чистої води, використовуючи процедуру,
розроблену Пустаї та ін.  (2008). Ця процедера поєднує повні
структурні фактори розсіяння і парціальні радіальні функції
розподілу, що слідують з молекулярної динаміки певної структурної
моделі. Якість узгодження значень функцій розподілу  O--O, O--H і H--H
з моделювання після їх перетворення до повних структурних факторів і
порівняння з експериментальними факторами є мірою для опису рівня
узгодження. Серед парціальних функцій розподілу, досліджених у цій
роботі, ті фактори, які відповідають моделі  SWM4-DP виявилися
найкращими, але лише незначно кращими від тих, що були отримані з
моделі BK3. Важливо відзначити, що усі три потенціали дають майже
еквівалентні радіальні функції розподілу, які узгоджуються з
експериментом, але невелика розбіжність спостерігається для функцій
розподілу киснів O--O.
\keywords нейтронна дифракція, парціальні радіальні функції
розподілу, метод реверсного Монте Карло
\end{abstract}


\end{document}